\documentstyle[prb,aps,graphicx]{revtex}  
\begin{document}
\draft

\def\ET {BEDT-TTF}                 
\def\ETX {({\ET})$_2$X}            
\def\KETX {$\kappa$-({\ET})$_2$X }  
\def\ETCl {({\ET})$_2$Cu[N(CN)]$_2$Cl}
\def\ETBr {({\ET})$_2$Cu[N(CN)]$_2$Br}
\def\k {{\bf k}}                   
\def\q {{\bf q}}                   
\newcounter{cms}

\title{Symmetrized mean-field description of magnetic instabilities in 
$\kappa$-({\ET})$_2$Cu[N(CN)]$_2$Y salts} 
 
\author{A.Painelli and A.Girlando}
\address{Dipartimento di Chimica Generale ed Inorganica, Chimica Analitica, 
Chimica Fisica, Universit\`a di Parma, 
Parco Area delle Scienze, I-43100, Parma, Italy} 
 
\author{A.Fortunelli} 
\address{Istituto di Chimica Quantistica ed Energetica Molecolare del CNR, 
v. V. Alfieri 1, I-56010, Ghezzano (PI), Italy} 
 
\date{\today} 
\maketitle 
 
\begin{abstract}
We present a novel and convenient mean-field method, and
apply it to study the metallic/antiferromagnetic interface of
$\kappa$-({\ET})$_2$Cu[N(CN)]$_2$Y organic superconductors (BEDT-TTF
is bis-ethylendithio-tetrathiafulvalene, Y=Cl,Br).
The method, which fully exploits the
crystal symmetry, allows one to obtain the mean-field solution
of the two-dimensional Hubbard model
for very large lattices (tipically $6 \times$10$^5$ sites),
yielding a reliable description
of the phase boundary in a wide region of the parameter space.
The metal/antiferromagnet transition appears to be
second order, except for a narrow region of the parameter
space, where the transition is very sharp and possibly
first order. The coexistence of  metallic and antiferromagnetic
properties is only observed for the transient state in the case
of smooth second order transitions. The relevance of the present
results to the complex experimental behavior of
centrosymmetric $\kappa$-({\ET})$_2$Cu[N(CN)]$_2$Y salts
is discussed.

\end{abstract} 
\twocolumn
 
\section{Introduction}

\begin{figure}
\begin{center}
\protect
\includegraphics[scale=0.8]{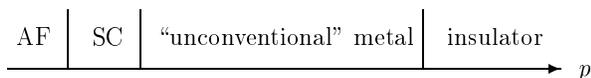}
\noindent
\caption{Universal zero temperature phase diagram for $\kappa$-phase
BEDT-TTF salts.}
\label{pd}
\end{center}
\end{figure}

The $\kappa$-phase {\ETX} salts exhibit a great variety of physical 
properties as a function of temperature, pressure, anion (X) 
substitution, deuteration, and even disorder in the ethylene 
end-groups. Superconducting (SC), antiferromagnetic (AF), 
metallic and insulating phases are observed.\cite{mckenzie1997,mori1999} 
Of particular interest is the AF/SC/metal borderline, which 
for $\kappa$-({\ET})$_2$Cu(NCS)$_2$ and $\kappa$-({\ET})$_2$Cu[N(CN)]$_2$Y 
(Y = Cl, Br; hereafter ET-Y family) occurs in a very narrow region 
of the temperature-pressure ($T,p$)
space. For the aforementioned compounds, a schematic zero temperature 
phase diagram can be drawn as 
shown in fig.~\ref{pd}.\cite{kino95,kanoda1997,mckenzie1998} 

The proximity of SC and AF phases, 
together with other experimental evidences, 
suggested a possible 
role of spin fluctuations in the superconductivity 
mechanism,\cite{mckenzie1998,schmalian1998} 
and prompted intensive theoretical investigation on the SC/AF  borderline.
\cite
{mckenzie1998,schmalian1998,ivanov1998,goddard,imamura1999,visentini99,seo} 
Since the early suggestion by Kino and Fukuyama,\cite{kino95,kino96}
mean-field (mf) approaches have often been adopted to investigate the SC/AF 
interface.\cite{goddard,imamura1999,seo}
Several previous mf treatments considered 
low-symmetry structures,\cite{kino95,goddard,kino96}
namely $\kappa$-({\ET})$_2$Cu(NCS)$_2$, with four inequivalent molecules
in the unit cell. The resulting numerical calculation is complex
and computationally very demanding, so that only fairly small lattices
have been considered, leading to large intrinsic uncertainties
on the estimated properties, particularly at the phase transition.
On the other hand, in the orthorhombic centrosymmetric
$\kappa$-phase crystals of the ET-Y family
all  molecules in the layer are equivalent.\cite{williams,willsci} 
The greater symmetry with respect to 
$\kappa$-(BEDT-TTF)$_2$Cu(NCS)$_2$ 
apparently does not lead
to significant differences in the physical behavior,
notably in the SC properties.
Indeed, ET-Br is a superconductor at ambient pressure, and
ET-Cl under moderate pressure presents
the highest $T_c$ observed in {\ETX} salts.\cite{mori1999}
The mf approach has been applied 
\cite{imamura1999,seo} also to these
more symmetric lattices to study the coexistence of charge
ordering instabilities. Including the order parameters relevant to charge 
instabilities  further increases the complexity of the 
calculation and leads to  large uncertainties in the transition region, 
as recognized by the authors of Ref. \onlinecite{imamura1999}.
We instead propose to fully exploit the higher symmetry of 
the centrosymmetric ET-Y structures to build a 
{\it symmetrized} mf approach 
for the superconducting metallic/magnetically ordered phase
transition.\cite{visentini99} 
By exploiting symmetry we are able
to work on lattices as large as $6 \times$10$^5$ sites,
and to explore a wide region of the 
parameter space, keeping numerical procedures and
finite-size effects under control. Moreover and most importantly,
we get a simple and complete
description of the electronic bands of these system, focusing only
on those effects which are directly connected with the 
interesting physics. 
The mf treatment maps the problem of interacting electrons
into an effective non-interacting Hamiltonian.
The comparison between reliable mf results and available
experimental data then allows us to safely define the intrinsic
limitations of effective one-electron pictures in describing
the physics of $\kappa$-phase salts.

Recently more refined approaches have been applied
to $\kappa$-phase salts.
For instance, fluctuation-exchange (FLEX)
\cite{kk,kondo} or third-order perturbation (PT) approach \cite{jujo}
have been applied to investigate the fluctuation mechanism
for superconductivity and to estimate the critical temperature.
However these approaches are only valid in the weak electronic
correlations regime, and have been applied to a simplified model
for the $\kappa$-phase layer, the so-called dimer
model.\cite{kino95,mckenzie1998,caufield94,visentini98}
The same model has also been adopted in the framework
of the dynamical mf\cite{mckenzie1998,voll} and within 
a renormalization group approach,\cite{tsai} which offer
complementary information with respect to ordinary mf techniques.
As we shall discuss in the following, the reliability of the
dimer model cannot be taken for granted in the whole
parameter space. Here we show that by exploting the high
symmetry of the ET-Y salts family one can get a picture
of the $\kappa$-phase layer that is computationally and
theoretically as simple as the dimer model, without
introducing any approximation.

In this paper we model the system in terms of a simple Hubbard,
$t-U$, Hamiltonian, but  the proposed procedure can be
easily extended to $t-U-V$ or $t-J$~~ Hamiltonians,
to investigate charge-ordering transitions,
whose possible coexistence with spin-order
has been recently suggested.\cite{mazumdar}
Moreover, the symmetry properties can be conveniently implemented
in more refined calculation schemes, to get
simpler and more reliable description 
of the physics of $\kappa$-phase salts.
The paper is organized as follows. The next section is devoted
to the description of the method. We then analyze the magnetic
instabilities of ET-Y salts,
and discuss the effects of the instabilities
on the band structure.
The difference between our symmetrized mf and the other
mf approaches is stressed, and the reliability of the dimer
model is shortly addressed. Finally, we
make connection with the experiment by discussing the pressure
dependence of the SC/AF interface in the ET-Cl, and by
making a comparison with ambient pressure ET-Br superconductor.

\section{The symmetrized mean field approach}

Consistently with experimental data
on centrosymmetric $\kappa$-phase salts,\cite{williams,willsci}
we consider a  unit cell with 4 equivalent molecular sites,
and do not allow for modification of the periodicity of the crystal structure 
at  the magnetic phase transition.
We adopt the $t-U$ Hubbard Hamiltonian to describe  Coulomb interactions 
giving rise to magnetic ordering: 
 
\begin{equation} 
H = \sum_{<i,j> \sigma} t_{ij} (a^{\dagger}_{i \sigma} a_{j \sigma} 
+ {\rm h.c.}) + \frac{U}{4} \sum_i n_i n_i - U \sum_i s_i s_i
\label{ht} 
\end{equation} 
where the indices run on the BEDT-TTF sites,
the first term accounts for the intersite hopping, 
and the other terms describe the on-site Coulomb repulsion. 
In eq.(\ref{ht}), $n_i$ is the usual site number operator $n_i
= n_{i \uparrow} + n_{i \downarrow}$, and
 $s_i = (n_{i \uparrow} - n_{i \downarrow})/2$ is the net magnetization
operator.

\begin{figure}
\begin{center}
\includegraphics* [scale=0.4,angle=0]{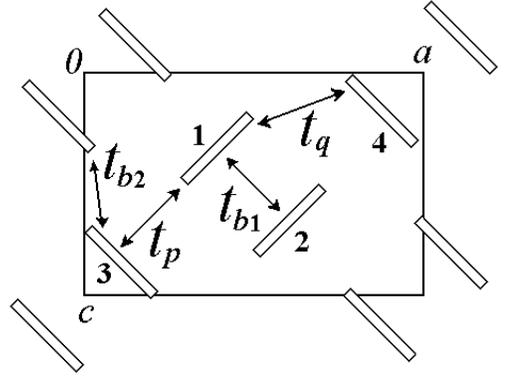}
\noindent
\caption{Schematic view of the $ac$ plane of centrosymmetric BEDT-TTF salts.}
\label{struct}
\end{center}
\end{figure}

In the mf approximation the many-body interaction is 
described by an  effective 
single particle interaction, where each 
particle feels  the other particles as a source of  
a  mf potential. 
Then  each product of two electronic operators $\hat A \hat B$ 
is approximated with an expression where only a single operator 
appears, the effect of the second operator being 
substituted by its ground-state expectation value. 
This approach gives reliable results when the 
fluctuations of the observables are small, although not zero as in
single-particle approaches. Mathematically: 

\begin{eqnarray} 
\hat A \hat B &  = & (\langle A \rangle + \widehat {\Delta A}) 
(\langle B \rangle + \widehat {\Delta B}) \cr
 & \approx &  \langle A \rangle \langle B \rangle 
+ \langle A \rangle  \widehat {\Delta B} 
+ \langle B \rangle  \widehat {\Delta A} 
\label{mf} 
\end{eqnarray} 
Thus in mf the two-particle Hubbard terms of eq. (\ref{ht}) become:
\begin{equation}
\frac{U}{4} \sum_i n_i n_i - U\sum_i s_i s_i~~\simeq~~\frac{U}{2}
\sum_i \langle n_i \rangle n_i - 2U \sum_i  \langle s_i \rangle  s_i
\label{Uterm}
\end{equation}
The equivalence of the four molecular sites imposes the constraint:
\begin{equation} 
\langle n_{i \uparrow} \rangle + \langle n_{i \downarrow}\rangle = 1.5 
\quad i = 1 \ldots 4 . 
\label{constraint} 
\end{equation} 
Therefore the first term on the r.h.s. of eq.~(\ref{Uterm}) is a constant, and
the relevant physics is described by the net magnetization term. 
We rewrite it by exploiting  symmetry,
and define within each unit cell the following four order parameters: 
\begin{eqnarray} 
\protect\label{salc} 
\eta_{AF1}~~~& = &  (s_1 + s_2 - s_3 - s_4) \nonumber \\ 
\eta_{AF2}~~~& = &  (s_1 - s_2 - s_3 + s_4) \nonumber \\ 
\eta_{AF3}~~~& = &  (s_1 - s_2 + s_3 - s_4) \\ 
\eta_{FM}~~~~& = &  (s_1 + s_2 + s_3 + s_4) \nonumber 
\end{eqnarray} 
or, in short: $\eta_\nu = \sum_i c_i^\nu s_i$, with
$\nu$ = AF1, AF2, AF3, FM. In these equations
$i$ counts the four BEDT-TTF sites within the unit cell, as indicated in
fig.~\ref{struct}. The order parameters $\eta_{AF1}$, 
$\eta_{AF2}$ and $\eta_{AF3}$ correspond to the three possible 
antiferromagnetic orderings; $\eta_{FM}$ describes the 
ferromagnetic phase. The net magnetization term in 
Eq.~(\ref{Uterm}) then becomes:
$ - 2U \sum_i  \langle s_i \rangle  s_i
= - (U/2) \sum_{j,\nu} \langle \eta_{\nu} \rangle 
\eta_{\nu}^{(j)}$ 
where $j$ runs over the unit cells. Since translational symmetry is 
not broken by the magnetic transition,  $\langle \eta_{\nu} \rangle$
is independent on $j$. The four magnetic phases have 
different symmetry, so that the four $\eta_{\nu}$ order parameters are 
orthonormal and can be investigated separately, leading to 
the symmetrized mf Hamiltonians: 
\begin{equation} 
H_\nu  =  - \sum_{<l,k>,\sigma} 
t_{lk}(a^{\dagger}_{l\sigma}a_{k\sigma} + h.c.) 
 - Y_\nu \sum_j \eta_\nu ^{(j)}
\label{hammf} 
\end{equation} 
where $t_{lk}$ are the hopping parameters, i.e., 
$t_{b_1}$, $t_{b_2}$, $t_{p}$, and $t_{q}$
defined in fig. \ref{struct}. 
For each symmetry, the effective
single particle potential $Y_\nu$, is related to the expectation 
value of the relevant order parameter by the self-consistency equation:
\begin{equation} 
Y_\nu   =  \frac{U}{2} \langle \eta_\nu \rangle 
\label{selfcons} 
\end{equation}

\noindent
By rewriting the  last term in eq.~(\ref{hammf}) 
in terms of the original $n_{i\uparrow}$, $n_{i\downarrow}$ operators,
one immediately recognizes that $H_\nu $ is the sum of two independent
tight-binding Hamiltonians, $H_{\nu \uparrow}$,  $H_{\nu \downarrow}$,
describing electrons with up and down spin, respectively.
The off-diagonal part of each one of these Hamiltonians 
is exactly the same as in the original tight-binding model, but
the mf treatment of on-site electron-electron interaction introduces a
diagonal contribution. Specifically, the diagonal elements of
$H_{\nu \uparrow}$ within each unit cell are:
\begin{equation} 
(H_{\nu \uparrow})_{ii} = -\frac {Y_\nu} {2} c_i^\nu 
\label{otto}
\end{equation} 
and $(H_{\nu \uparrow})_{ii} =-(H_{\nu \downarrow})_{ii}$.
The two tight-binding problems described by $H_{\nu \uparrow}$
and  $H_{\nu \downarrow}$ are easily diagonalized
for different $Y_\nu$ values on very large lattices.
In our approach imposing the self-consistency relation
on $U$ simply implies the ratioing of $Y_\nu$ and $\langle\eta_\nu\rangle$,
at variance with the lengthy and memory consuming iteration
steps required by a multi-parameter mf calculation.
\cite{kino95,goddard} This is very important in keeping
the numerical procedure under control and allows us
to work with very large lattices, tipically up
to 6$\times$10$^5$ sites.
Such large lattices, one order of magnitude larger
than the largest lattice in Ref.~\onlinecite{goddard},
are diagonalized with no effort on a Digital Alpha
255 workstation equipped with 64 MB RAM.
As we will discuss below, working on large lattices is very important to 
get an accurate description of the early stages of the phase transition, 
and then to get reliable information on the nature of the transition itself. 

The diagonalization of $H_{\nu \uparrow}$,   $H_{\nu \downarrow}$
immediately defines the band structures for up and down spins. 
In the case of the FM instability, all the $c^{FM}_\nu$ in eq.~\ref{otto}
are equal to 1, so that, apart from a rigid shift of the 
energies by $-(+)Y_{FM}/2 = -(+) U\langle \eta_{FM} \rangle /4$
for up (down) spins, the eigenstates are exactly the same as in the
non-interacting case. Therefore, the originally degenerate bands for up and
down spins are split by $Y_{FM}=U\langle \eta_{FM} \rangle$.
The Fermi level is fixed by the conservation of the total number of
electrons, leading to unbalanced up and down spin population.
If, without loss of generality, we consider positive 
$\langle \eta_{FM}\rangle$, we end up with lower energies for up
spins and then with a ferromagnetic state characterized by
larger population of up than down spins.

In the case of AF order, instead, finite $Y_{\nu}$ deform the
original bands of the non-interacting system, due to the appearance of 
relevant diagonal terms in the real space Hamiltonian (eq.~(\ref{otto})).
In this case,  the eigenvalues for up and down spins are exactly the same,
and the bands for the two spins stay exactly degenerate as in the 
non-interacting case, but the distribution of the two spin species 
is different on the sites, with a larger number of up spins residing on 
sites with negative $c^\nu_i$ coefficients
($\langle \eta_{AF} \rangle >0$).

\section{Results} 

Table \ref{tabella} summarizes the $t$'s
obtained from the available structural data 
of ET-Cl and ET-Br salts.\cite{geiser1991,HPstruct}
All $t$'s have been obtained from 
extended H\"uckel (EH) calculations on the pairs of BEDT-TTF molecules
corresponding to the four interactions depicted in fig.~\ref{struct}.
Specifically, they are evaluated as half of the splitting of the HOMO 
energy in each pair.
It is well known that the values of the hopping integrals
show large differences, depending on the method
adopted for their estimate.\cite{mckenzie2000}
Therefore, the $t$'s estimated for each structure
and the resulting $U_c$ have not to be assigned too much
confidence. However, comparing results obtained with the
same procedure on different structures is certainly informative.
We have adopted EH estimates of 
$t$'s since they compare well 
with available {\it ab initio} results.\cite{fortunelli}

\begin{figure}
\begin{center}
\includegraphics* [scale=0.45,angle=0]{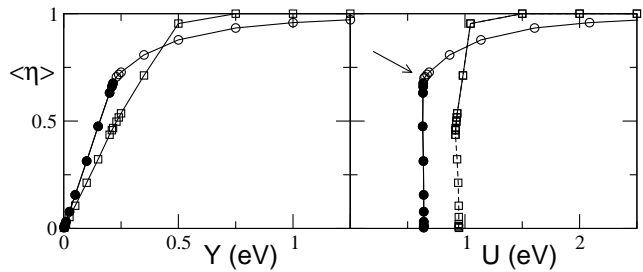}
\vskip 5mm
\caption{ET-Cl at $T$ = 127 K, ambient $p$:
$t$'s from the first row of Table \protect\ref{tabella}.
The order parameter for FM instability (squares) and AF instability
of type 1 (circles) vs the effective mf potential $Y$, in the
left panel, and vs $U$ in the right panel. Filled circles mark points
introduced in the linear regression of the $\eta$ vs. $Y$
to determine $U_c$ (see text), the arrow marks the point where the
conduction gap opens, and whose band structure is shown in fig.~
\protect\ref{band}, bottom panel.}
\label{eta}
\end{center}
\end{figure}

Fig.~\ref{eta} reports the $Y_\nu$ and $U$ dependence of
$\langle \eta_{AF1} \rangle$ and 
$\langle \eta_{FM} \rangle$ order parameters, as obtained for the $t$'s 
relevant to ET-Cl at 127 K (first row in Table \ref{tabella}).
$\langle \eta_{AF2} \rangle$ and $\langle \eta_{AF3} \rangle$ 
curves are not shown since the corresponding instabilities occur at
$U$ larger than $\sim$ 1 eV,\cite{visentini99} and are
not relevant to our discussion. Indeed, even the ferromagnetic
instability occurs at $U$ higher than that for AF instabilities, and,
in this respect, it is irrelevant from the physical point of view. 
However, the different behavior of
$\langle \eta_{AF1} \rangle$ and $\langle \eta_{FM} \rangle$
in the right panel of fig.~\ref{eta} deserves some comments.

Based on the two standard stability conditions: $\partial E/\partial \eta =0$
and  $\partial ^2 E/\partial \eta ^2 >0$, with $E$ representing the
expectation value of the working Hamiltonian, it is easy to prove
that stable states for our system 
correspond to points with positive slope in the $\eta (U)$ curves.\cite{PGPRB}
The negative slope region 
in the $\eta_{FM}(U)$ curve (marked by 
a dotted line in the right panel of fig. \ref{eta})
then corresponds to unstable states,
i.e. states that cannot be reached by our physical system. 
Thus the ferromagnetic instability corresponds to a first order phase
transition, characterized by a discontinuous jump of the order parameter
at the transition, located at $U_c \sim 0.945$ eV.
The small region around $U_c$ where the $\eta_{FM}(U)$ is non-single-valued
corresponds to the hysteresis region, where two stable states coexist.

The behavior of $\langle \eta_{AF1} \rangle$
is different, with $\eta_{AF1}(U)$ having infinite
slope at $U_c \sim 0.64$ eV. The infinite slope is a direct consequence 
of a strictly linear $\eta_{AF1}(Y_{AF1})$ dependence in a fairly large 
region around the origin. In fig.~\ref{eta}, left panel, the filled
circles show the
points that fall on a single straight line,
$\langle \eta_{AF1} \rangle =\chi Y_{AF1}$,
with a squared correlation coefficient larger than 0.99998.
By applying the self-consistency condition in eq.~(\ref{selfcons}),
one immediately gets steeply increasing $\eta_{AF1}$ values at a fixed
$U=U_c=2/\chi = 0.639$ eV. The coefficient
$\chi =\partial ^2 E/\partial Y _{AF1}^2$ represents the susceptibility of 
electronic system to the $Y_{AF1}$ perturbation: the critical $U$ is thus
related to the inverse of the electronic susceptibility.

Extracting $U_c$ from the slope of the $\langle \eta_{AF1}\rangle$ vs 
$Y_{AF1}$ curve is a much safer procedure than searching for the
minimum $U$ where finite $\langle \eta_{AF1}\rangle$ appears.
The calculated $\langle \eta_{AF1}\rangle$ values are affected by finite
uncertainties, with a minimum intrinsic uncertainty given by the inverse of the
number of unit cells. Since  $\langle \eta_{AF1}\rangle$ 
enters the Hamiltonian 
matrix as a multiplicative factor for $U$, the uncertainty in
$\langle \eta_{AF1}\rangle$ implies an uncertainty in $U$, with
$\delta U/U = \delta\langle \eta_{AF1}\rangle/\langle \eta_{AF1}\rangle$.
Therefore,  at small $\langle \eta_{AF1}\rangle$ the relative
uncertainty on $U$ can be very large. 
This is by no means accidental, but reflects the intrinsic limitation of 
investigating phase transitions through finite size calculations.
At the transition in fact the correlation length of the fluctuations
in the order parameters are in principle infinite, so that calculations
on finite lattices lead to large errors.

\begin{figure}
\begin{center}
\protect
\includegraphics* [scale=0.46,angle=0]{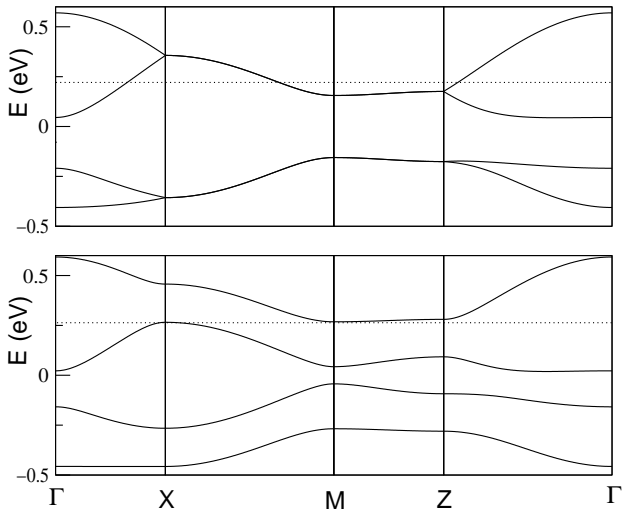}
\vskip 5mm
\caption{Band structure for ETCl, same parameters as in fig.~\ref{eta}. 
The upper panel refers to the non-interacting system (or equivalently 
to the system before the transition), the bottom panel corresponds
to the point marked by an arrow in fig.\protect\ref{eta}, $Y = 0.225$ eV.
The dotted line marks the Fermi energy.}
\label{band}
\end{center}
\end{figure}

\begin{figure}
\begin{center}
\includegraphics* [scale=0.7,angle=0]{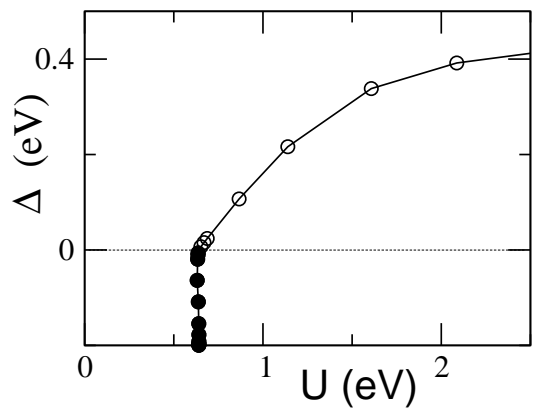}
\vskip 3mm 
\caption{The conductivity gap (see text) vs $U$, same parameters as
in fig.\protect\ref{eta}.}
\label{gap}
\end{center}
\end{figure}

It is interesting to investigate the evolution of the electronic bands
along the AF transition. Fig. \ref{band} reports the band-structure
calculated for the non-interacting metallic system,
and for a system located just where the transition goes
to completion, i.e. the point marked by an arrow on fig.~\ref{eta}. 
The two conduction bands, that are partly overlapped in the
metallic system, are split apart in the AF phase,
opening a gap and then leading to insulating behavior.
Fig.~\ref{gap} reports the $U$-dependence of the energy difference
between two extreme points in the two conduction bands 
(specifically between M-point in the upper band and X-point in the
lowest conduction band) to measure the conductivity gap, $\Delta$. Negative
$\Delta$ implies overlapping bands and then metallic behavior, 
positive $\Delta$ measures the semiconducting gap. 

In summary, for the $t$'s in the first row of Table \ref{tabella},
relevant to ET-Cl
at ambient pressure and $T$ = 127 K, we observe a
fairly sharp transition, at $U_c$=0.639 eV, from
a paramagnetic metal to an antiferromagnetic insulator, as shown by
the semiconducting gap that opens up right at the transition (fig. \ref{gap}).
In our approach the metallic phase includes the superconducting state, since
our Hamiltonian does not account for SC coupling. The critical $U$ is similar
to available experimental\cite{mckenzie1998,Uvalue}
and theoretical\cite{fortunelli}
estimates of the effective $U$ in
BEDT-TTF salts, $U \sim$ 0.5 - 1.0  eV.
Therefore ET-Cl is just located at the metal/AF interface, in  agreement
with several experimental observations (see below).
Again, in agreement with experiment and also with predictions of previous mf
calculations,\cite{kino95,goddard,kino96}
the AF phase is characterized by parallel spins
residing on the 1-2 dimer (fig.\ref{struct}), as
could also been inferred from simple
arguments based on the dimer picture.\cite{kino95}

\begin{figure}
\begin{center} 
\includegraphics* [scale=0.7,angle=0]{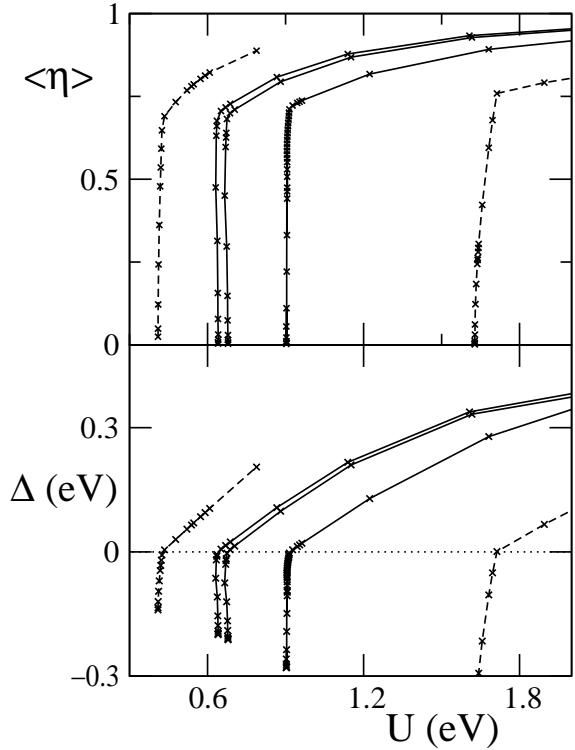}
\caption{The AF order parameters and the conductivity gap vs $U$.
Continuous lines refer to $t$-values in Table \ref{tabella}, with
$p$ increasing from left to right. Dashed lines refer to nominal
pressures $p = -$33 kbar, at left, and $p$ = 95 kbar, at right (see text).}
\label{etagap}
\end{center}
\end{figure}

Having developed a simple and efficient method to solve the mf
problem for the ET-Y family, we can now play around with parameters
trying to gain some
information about the rich phase diagram of these systems. 
In fig.~\ref{etagap} the  continuous lines show the $U$-dependence of 
$\langle \eta_{AF1}\rangle$ and $\Delta$, calculated for 
the available $t$'s relevant to ET-Cl
at  $p$ = 3 and 27 kbar (Table \ref{tabella}).
The critical $U$ increases with $p$
from $\sim$ 0.64 eV at $T$ = 127 K, ambient $p$,
to  $\sim$ 0.68 eV and  $\sim$ 0.91
eV, at ambient $T$ and $p$ = 3, 27 kbar, respectively. 
The increase of $U_c$ corresponds to a stabilization of the metallic phase,
and can justify the appearance of SC in ET-Cl under pressure,
as we shall discuss in more detail in next Section.

In the scale of fig.~\ref{etagap} the curve relevant to $p$ = 3 kbar
shows a very narrow region with a negative slope. However, the width of this
region is only 2-3 times the numerical uncertainty on $U$, so that we
cannot  make any strong statement about observing a discontinuous, 
first order transition. In any case, the coexistence region, i.e. the
hysteresis region for this transition, if present, would
be so small to be irrelevant for any practical purpose. The region of negative
slope disappears at $p$ = 27 kbar, where the transition looks
smoother, with possibly a finite positive slope.
Once more  the effect is tiny  and  hardly  disentangled
from numerical uncertainties.

To get some clearer feeling about  the role of pressure,
we have linearly extrapolated 
the $t$-estimates available at $p$ = 3 and 27 kbar to higher pressures.
In fig.~\ref{etagap} the rightmost dashed line shows the corresponding 
evolution of the order parameter for a nominal $p$ = 95 kbar,
where $t_q$ extrapolates to zero.
The smoothing of the transition is now evident: in this case
$\langle \eta_{AF1}\rangle (U)$ clearly has a well-defined positive slope. 
An interesting observation is that the conduction gap (lower panel)
closes not at the very beginning of the transition, but only when the
transition comes to completeness. This corresponds to the appearance
of a region of stability for a phase with simultaneously AF distortion
and residual metallic behavior. In other words, when the transition
is continuous, our data suggest the presence of an antiferromagnetic
metal phase, similar to that discussed at
length in Refs.~\onlinecite{kino95,goddard,kino96}.
However, the extent of this phase is very narrow, so that it represents
at most a transient phase.

Searching for some evidence of discontinuous phase transition,
we have also ``released the pressure'', by extrapolating the $t$'s
estimates at $p$ = 3 and 27 kbar down to a nominal
pressure of $-$33 kbar. Surprisingly, the transition is continuous
again, and 
smoother than at ambient $p$. Therefore, we find no clear evidence for 
discontinuous transitions, except possibly in a very narrow region
in the parameter space corresponding to ET-Cl at $p$ = 3 kbar.
In general, the observation of very narrow regions of discontinuity and/or
AF-metallic coexistence is strongly affected by numerical uncertainties
and/or finite-size effects.\cite{finite-size} These transient states, being
intrinsically unstable, cannot be associated with
physically significant states.
On the other hand, their presence for particular values of the parameters
signals an intrinsic instability of the system to external perturbations,
such as those eventually leading to superconductivity.

\section{Discussion}

Overall our results agree with previous mf calculations on 
$\kappa$-(\ET)$_2$Cu(NCS)$_2$,\cite{kino95,goddard,kino96}
describing the transition from a paramagnetic 
metal to an AF insulator occurring at $U_c \sim$ 0.6-0.8 eV 
(the precise value of course depends on the choice of the $t$'s).
However, some details on the evolution of the order parameter and on the
opening of the semiconducting gap are different.
In the recursive approach to the solution of the mf problem, 
adopted so far in the literature, 
the only viable procedure to estimate $U_c$ relies 
on searching for the minimum $U$ where finite $\langle \eta_{AF1} \rangle$
appears. As discussed above, this procedure
leads to large uncertainties in $U_c$, that have to be properly
accounted for in the analysis of numerical results.
Kino and Fukuyama\cite{kino95,kino96} use very small lattices (N=3600),
corresponding to an intrinsic uncertainty in 
$\langle \eta_{AF1} \rangle$ of at least $4 \times 10^{-4}$. 
In Ref. \onlinecite{kino95} the onset of AF is estimated to occur at
$U_{c2}$, with $\langle \eta_{AF1}\rangle \sim 0.02$.
This small value for the order parameter
implies a minimum uncertainty in $U_{c2}$ of $\sim$ 0.02 eV.
Then the two transition points observed by Kino and
Fukuyama, $U_{c1}$ =0.762 eV and  $U_{c2}$ = 0.758 eV, coincide within
numerical accuracy. In the lack of additional information their data
are consistent with a single transition, as we find for the $t$'s relevant
to ET-Cl at $T$ = 127 K or at $p$ = 3 kbar (fig.~\ref{etagap})
The presence of an antiferromagnetic metallic phase is then questionable.
Similar problems occur in the interpretation of data in Ref.
\onlinecite{kino96}. Here a two-transition scenario is proposed at low $p$,
involving a continuous
transition from a paramagnetic metal to an antiferromagnetic metal
(finite $\eta_{AF1}$ and negative $\Delta$), immediately followed by
a first order transition to an insulating state. This complex scenario,
that we were unable to reproduce in our large lattice for any choice of the
parameter set, is probably either a finite-size effect or a numerical
artifact. One must also keep in mind the possibility it represents a
characteristic feature of $\kappa$(\ET)$_2$Cu(NCS)$_2$, due
to its  lower symmetry. However, since
it does not appear in the more symmetric ET-Cl phase at any pressure, it is
irrelevant as far as SC is concerned.

In Ref. \onlinecite{goddard} 
the numerical uncertainty in $\langle n_{i\uparrow}\rangle$, 
fixed by the authors at 0.001, propagates to 
give  $\delta \langle \eta_{AF1} \rangle \sim $ 0.0014. As a consequence, 
the estimate for $U_a$, i.e. the critical $U$ for the appearance of 
AF order, obtained for $\langle \eta_{AF1} \rangle = 0.01$, 
is affected by a large
uncertainty: $U_a =0.7 \pm 0.1 eV$.
More precise estimates are obtained for larger 
$\langle \eta_{AF1} \rangle$, e.g.
$U_c =0.685$ is essentially constant for  $\langle \eta_{AF1} \rangle $ =
0.037, 0.499, 0.582, representing a good estimate for the critical
$U$ where AF order appears and, at the same time, the electronic orbits close.
The proposed estimate of the critical $U$ for the opening of the 
semiconducting gap, $U_i =0.699\pm 0.001$ is 
different from $U_c$, again suggesting the presence of an intermediate phase
with both metallic and AF character.
Quite in agreement with our results at
large $p$, the metallic antiferromagnetic phase is a marginal phase
that only survives in a very narrow transient regime.
Indeed, as pointed out in Ref.~\onlinecite{singletonrev} (Section 3.4.5),
such a phase would imply a weak AF order and reconstruction of the
Fermi surface, which however have not been experimentally observed.
Our approach thus proves useful in excluding on a purely theoretical basis
the spurious complexities in the phase diagram due to finite-size effects
and/or numerical uncertainties, and should be particularly convenient
when extended to describe charge ordering instabilities together
with magnetic instabilities.

Several papers discuss $\kappa$-phase BEDT-TTF salts within the
the dimer approximation.\cite{kino95,mckenzie1998,caufield94,visentini98}
Basically, the tight-binding Hamiltonian for the four
frontier molecular orbitals in the unit cell is rewritten
in terms of the bonding and antibonding orbitals
of the $t_{b1}$-dimers. Since  $t_{b1}$ is at least twice
as large as the other hopping integrals, the interactions between 
bonding and antibonding orbitals are neglected, and the original 
four-bands problem reduces to a two-band problem. In the resulting lattice
each site has four nearest-neighbor sites, interacting through
$(t_p + t_q)/2$, and two next nearest-neighbors, interacting with $t_{b2}/2$.
For the parameters relevant to $k$-phase salts, the bands calculated
within the dimer model compare favorably with those obtained in the 
four-band calculation, confirming the validity of the dimer model 
approximation at least for the non-interacting case.\cite{visentini98}
The dimer-model lattice is simple, but still shows interesting physics.
In fact, by  varying the $(t_p + t_q)/t_{b2}$ ratio, it interpolates between
a square lattice and a collection of 1D chains.\cite{tsai} Whereas it is suggestive to
relate the variegated behavior obtained from such a model to the 
variety of observed properties for $\kappa$-phase salts, 
some caution is in order. Just as an example, consider the case 
$t_{b2} \rightarrow 0$, where the dimer lattice reduces to a half-filled
square lattice with perfect nesting.
As it is well-known, the critical $U$ 
for the antiferromagnetic instability goes to zero in this limit, as
also confirmed by mf calculations.\cite{visentini99} 
Instead, a mf calculation for the same parameters as in fig. 3,
but $t_{b2} = 0$, yields a continuous transition to the AF phase with
a finite and fairly large $U_c \sim 0.57$ eV. This
qualitatively different behavior can be easily 
rationalized: The small interactions between bonding and 
antibonding orbitals are large enough to break the commensurability of 
the simple dimer model at $t_{b2} = 0$.\cite{visentini99,tsai}
A word of caution is also necessary when introducing
electron correlations in the
dimer model. Indeed, starting from a Hubbard Hamiltonian for the
four-molecules layer, the resulting effective $U_{dim}$ for the dimer
model is related to both $U$ and $t_{b1}$, according to a relation
first proposed in Ref.~\onlinecite{kino95} and rediscussed and
extended in Ref.~\onlinecite{mckenzie1998}.
For the commonly accepted value of $U \sim 4t_{b1}$, $U_{dim}$
is of the order of $t_{b1}$.\cite{mckenzie1998}
Thus the applicability of FLEX and perturbartive approaches\cite{kk,kondo,jujo}
to the dimer model becomes questionable, since
these approaches work well in the limit
$U_{dim} \sim t_{b1} \ll (t_p + t_q)/2, t_{b2}/2$, where
the dimer model itself breaks down.

We now relate our results to the experimental observations
relevant to the ET-Y family. As mentioned above,
several evidences indicates that the ET-Y salts
are just at the AF/SC borderline. At ambient pressure the ET-Cl
salt is a Mott antiferromagnet, with a magnetic moment amplitude
of 0.45 $\mu _B$, \cite{miyagawa2000} that compares favorably 
with the present and previous \cite{kino95,goddard,kino96}
estimates of the magnetic order parameter.
By applying pressure above 300 bar, ET-Cl shows a
transition to complete superconductivity
at about 12 K. At lower pressures, reentrant and partial
superconductivity, with residual sample resistance, have been observed.
\cite{ito1996} The fully deuterated ET-Cl ($d_8$-ET-Cl) exhibits
analogous behavior, only requiring a slightly higher pressure (440 bar)
to reach superconductivity.\cite{ito1996} ET-Br  is superconducting
at ambient pressure ($T_c \sim$ 11 K) but the attainment of the
superconducting phase is affected both by the cooling rate and
by the deuteration of the sample.\cite{taniguchi2000}
It has been shown that by keeping constant the cooling rate at a
sufficiently low value, the partially deuterated $d_2$- and $d_4$-ET-Br
salts are superconducting at practically the same $T_c$ as the
undeuterated sample. The $d_6$- salt, on the other hand, exhibits
a complicated behavior attributed to the competition between 
superconducting and insulating phase.\cite{taniguchi2000}
Finally, the fully deuterated $d_8$-ET-Br is an antiferromagnetic
insulator (magnetic moment: 0.3 $\mu_B$),\cite{miyagawa2000}
and under pressure has a behavior similar to ET-Cl, reaching
complete superconductivity just above 60 bars.\cite{ito2000}
For the sake of completeness, we mention that the ET-I salt
is not superconducting, even when pressures up to 5 Kbars
are applied.\cite{geiser1991} This kind of behavior has been
ascribed to disorder. However, this compound is
the least investigated in the ET-Y family,
and we shall not consider it here.

Rather obviously, a mf approach is inadequate to describe
phenomena related to non-equilibrium states, disorder and/or
sample inhomogeneity, like cooling rate effects and
reentrant superconductivity. We therefore focus here
on the complete-SC/AF crossover affected by pressure
and/or by isotopic substitution.
The universal phase diagram in fig.~\ref{pd} presents the parameter
``pressure'' as the abscissa. It has been used to explain the
differences induced by the Cl-Br substitution or deuteration in ET-Y
salts, and similar effects.
In particular, the smaller radius of Cl$^-$ with respect to Br$^-$
implies a reduced effective pressure\cite{mckenzie1997}
in ET-Cl with respect to ET-Br
(see Ref.\onlinecite{mori1999} for a tentative numerical assessment of this
effect). A similar effect of reduced pressure can be associated to
deuteration, which corresponds to smaller end-group excursions
around their equilibrium values.
From the values of the corresponding $T_c$'s,
we can empirically associate an increase in $p$ of $\sim$ 380 bar
for the Cl-Br substitution, and a decrease of $\sim$ 140 bar for deuteration.
A rationalization of these tiny effects is fairly difficult.
We shall examine below whether our mf results help in this respect.

\begin{figure}
\begin{center}
\includegraphics* [scale=0.48]{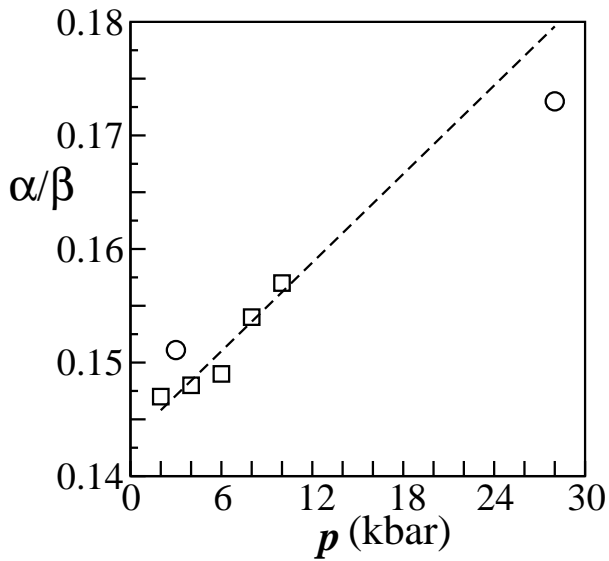}
\caption{Pressure dependence of the ratio between the areas
of $\alpha$ and $\beta$ orbits in the $ac$ plane Fermi surface.
The squares refer to the experimental data
(Ref.~\protect\onlinecite{kartsovnik}), with the dashed line
representing the best linear fit. Circles are the calculated values.}
\label{alphabeta}
\end{center}
\end{figure}

We first focus on ET-Cl. The hopping
integrals in the $ac$ plane, calculated for the known crystal structures
at ambient $p$ and 127 K,\cite {geiser1991}
and at 3 and 28 kbar (ambient $T$)\cite {HPstruct}
are reported in Table \ref{tabella}.
To make a first comparison with experiment, we have evaluated
the areas of the $\alpha$ and $\beta$ orbits from the $ac$ plane
Fermi surface calculated in the tight binding approximation.
The results relevant to the metallic phase are compared
in fig. \ref{alphabeta} with the measured areas from Shubnikov-de Haas
(SdH) experiments performed at several pressures in the 2-10 kbar
range at the liquid Helium temperature.\cite{kartsovnik}
In order  to renormalize the effects associated 
to the global volume contraction due to the 
different temperatures of the SdH and structural
data, the $\alpha/\beta$ ratio has been reported as a function of pressure. 
In fact, the $\beta$ area is equal to the area of the 
Brillouin zone for the 3/4-filled system.
The 3 kbar $\alpha/\beta$ nicely fits the experiment, and a straight
line through the experimental points extrapolates near
to the calculated 28 kbar point. We notice that the crystallographic
axes ratio $c/a$ is practically unchanged with pressure.\cite{HPstruct}
This observation sheds doubts on the possibility
of adopting $c/a$ as a rough estimate of $\alpha/\beta$,
as suggested by Ref.~\onlinecite{mori1999}.

Table \ref{tabella}, rightmost column,
reports the ET-Cl $U_c$ values calculated
by our mf approach.
A word of warning is necessary when comparing data obtained at
different temperatures, since it has been observed\cite{watanabe1999}
that the values of the hopping parameters change correspondingly.
However, this effect is not very pronounced in ET-Y
family,\cite{mori1999,watanabe1999}
and we shall neglect it in the following.
We notice that $U_c$ increases monotonously with $p$, thus
accounting for the pressure driven superconductivity transition in
terms of an increase of the critical value needed to reach the AF phase.
In this respect, $U_c$ seems to be a good ``indicator'' of the
effective pressure of fig.~\ref{pd}.
Other previously suggested indicators, like the $t_{b1}/t_p$
ratio,\cite{kino95,kino96} or the $c/a$ ratio,\cite{mori1999}
seem to work less satisfactorily in this case: $t_{b1}/t_p$
does not increase monotonously with $p$ (Table \ref{tabella}),
and, as noted above, $c/a$ is practically constant.

We now turn attention to ET-Br.
When comparing the ambient pressure, 127 K hopping integrals
of ET-Cl with the corresponding ones of ET-Br, one finds
small differences, and we get for the two systems 
virtually identical phase transitions, occurring at basically the same 
$U_c$  (Table \ref{tabella}). Therefore, the different ground state of ET-Cl
(AF) and ET-Br (SC) at ambient pressure cannot be understood
in terms of a difference in $U_c$.
One might think that the actual effective $U$ is different
in the two types of salts,
being smaller in ET-Br due a larger screening of the
intersite Coulomb potential from the more polarizable Br
anions. However, this kind of qualitative explanation is 
not corroborated by the numerical values of the anion polarizabilities
obtained from {\it ab initio} calculations,\cite{fortu}
and is difficult to reconcile with the observation
of an AF state for $d_8$-ET-Br at ambient pressure.
We could not calculate the hopping integrals
in this case, as the atomic coordinates are not available in
the literature. We have used the $t$'s calculated in
Ref.~\onlinecite{watanabe1999} for both ET-Br and $d_8$-ET-Br
at 127 K (properly rescaled since the method of calulation is different
from ours). We do not find significant difference between the
$U_c$'s of the two compounds.

As we already pointed out, $c/a$ or $t_{b1}/t_p$ are not good
indicators of the properties of $\kappa$-phase salts, and
cannot be chosen as the $x$-axis parameter in a $universal$
phase diagram like that reported in fig.~\ref{pd}.
Both the ratio of $\alpha$ and $\beta$ orbits and $U_c$ work
satisfactorily as far as the $p$-dependence of ET-Cl properties
is concerned. However, both fail if applied to rationalize
the different behavior of ET-Cl and ET-Br and/or the effects
due to deuteration. It is important to underline that
both $\alpha/\beta$ and $U_c$ are "single particle"
parameters, in the sense that they are fully determined by the
band structure, i.e., the $t$'s values. Investigating the
band structure of $\kappa$-phase salts offers useful information
to rationalize their behavior, but this information is not enough,
and the role of interactions beyond single-particle picture has to be
invoked to understand the behavior of systems near the AF/SC interface.
The failure of simple band-structure treatments for $\kappa$-phase
salts has been recently suggested based on the $p$-dependence
of cyclotron effective masses with pressure,\cite{mckenzie2000}
and can also be recognized from high resolution mesurements
of thermal expansion coefficients.\cite{muller}
Residual electronic correlations, disorder induced localization
effects, electron-phonon coupling 
and interlayer effects all can play an important role, particularly
at the AF/SC interface. More theoretical and experimental work
is in order to settle the relative importance of these effects.

\acknowledgments

This work has been supported by the Italian National
Research Council (CNR) within its ``Progetto Finalizzato
Materiali Speciali per tecnologie Avanzate II'', and the
Ministry of University and of Scientific and Technological
Research (MURST).



\onecolumn

\vskip 40mm

\begin{table}
\mediumtext
\protect
\caption{Hopping integrals and critical $U$ for ET-Cl and ET-Br salts.
All parameters in eV.}
\begin{tabular}{lrrrrr}
  & $t_{b1}$~~ & $t_{b2}$~~ & $t_p$~~~ & $t_q$~~~ & $U_c$ ~~~~~~ \\
\hline
ET-Cl, amb. $p$, $T$ =127 K ~ & 0.2315 & 0.0760 & 0.0901 & 0.0410 & 0.639$\pm$ 0.001 \\
ET-Cl, $p$=3 Kbar, amb. $T$ ~ & 0.2239 & 0.0851 & 0.0844 & 0.0517 & 0.676$\pm$ 0.005  \\
ET-Cl, $p$=27Kbar, amb. $T$ ~ & 0.2770 & 0.0935 & 0.1400 & 0.0380 & 0.906$\pm$ 0.005  \\
& & & & & \\
ET-Br, amb. $p$, $T$ =127 K ~ & 0.2244 & 0.0712 & 0.0936 & 0.0396 & 0.636$\pm$ 0.001 \\
\end{tabular}
\widetext
\label{tabella}
\end{table}

\end{document}